\documentclass[11pt,fleqn]{iopart}
\usepackage[utf8x]{inputenc}
\usepackage[T1]{fontenc}
\usepackage{fouriernc}
\expandafter\let\csname equation*\endcsname=\relax
\expandafter\let\csname endequation*\endcsname=\relax
\usepackage{amsmath,amssymb,amsfonts,amsthm}
\usepackage{graphicx}
\usepackage{hyperref}
\usepackage{units}

\numberwithin{equation}{section}
\def \nn{\nonumber}
\def \nl{\nn\\}
\def \ll{\left|}
\def \rr{\right|}
\def \lle{\left\langle}
\def \rre{\right\rangle}
\def \rr{\right|}

\newcommand{\bra}[1]{\lle #1 \rr}
\newcommand{\ket}[1]{\ll #1 \rre}
\newcommand{\brkt}[2]{\lle #1 | #2\rre}

\begin{document}
\title[Algebraic interpretation of $q$-Meixner polynomials]{An algebraic interpretation of the $q$-Meixner polynomials}
 \author{Julien Gaboriaud}
 \ead{julien.gaboriaud@umontreal.ca}
 \address{Centre de Recherches Math\'ematiques, Universit\'e de Montr\'eal, Montr\'eal, QC, Canada}
 \author{Luc Vinet}
 \ead{luc.vinet@umontreal.ca}
 \address{Centre de Recherches Math\'ematiques, Universit\'e de Montr\'eal, Montr\'eal, QC, Canada}
\begin{abstract}
 An algebraic interpretation of the $q$-Meixner polynomials is obtained. It is based on representations of $\mathcal{U}_q(\mathfrak{su}(1,1))$ on $q$-oscillator states with the polynomials appearing as matrix elements of unitary $q$-pseudorotation operators. These operators are built from $q$-exponentials of the $\mathcal{U}_q(\mathfrak{su}(1,1))$ generators. The orthogonality, recurrence relation, difference equation, and other properties of the $q$-Mexiner polynomials are systematically obtained in the proposed framework.
\end{abstract}
\vspace{2pc}
\hspace{2em}{\it\footnotesize Keywords}{\footnotesize: $q$-Meixner polynomials, $\mathcal{U}_q\mathfrak{su}(1,1)$ quantum algebra, $q$-oscillators}
\newpage

\section*{Introduction}
\label{sec:intro}
The purpose of this paper is to introduce an algebraic interpretation of the univariate $q$-Meixner polynomials. The $q$-Meixner polynomials, $\mathcal{M}_n\left(q^{-x};b,c;q\right)$ are defined as follows \cite{rf2010_Koekoek_Lesky_Swarttouw_hypergeometric}
\begin{align}
\label{eqHyperMeix}
\mathcal{M}_n\left(q^{-x};b,c;q\right) = {}_{2}\phi_{1}\left({q^{-n},q^{-x}} \atop {bq} \middle| q;-\frac{q^{n+1}}{c} \right)~,
\end{align}
with the basic hypergeometric series given by
\begin{align}
\label{eqHyperDef}
{}_{r}\phi_{s}\left({a_1,\ldots,a_r} \atop {b_1,\ldots,b_s} \middle| q;z \right) =\sum_{n=0}^{\infty}\frac{(a_1;q)_n\cdots(a_r;q)_n}{(q;q)_n(b_1;q)_n\ldots(b_s;q)_n}\left[(-1)^n q^{\binom{n}{2}}\right]^{1+s-r} z^n~,
\end{align}
where $\binom{n}{2}$ is the standard binomial coefficient, and $(a;q)_n = (1-a)(1-aq)\cdots(1-aq^{n-1})$ stands for the $q$-Pochhammer symbol.

These polynomials were introduced by Meixner \cite{rf1934_Meixner_polyn} in 1934 as the ones orthogonal with respect to the negative binomial distribution. Their connection with $\mathfrak{su}(1,1)$ is well established (see for instance \cite{rf1982_Basu_Wolf_irrepsSL2R, rf1991_Vilenkin_Klimyk_specialfct, rf1986_Granovskii_Zhedanov_su11Meixner, rf1993_Floreanini_LeTourneux_Vinet_quantmechpolyndiscrete}). Some 40 years later, Griffiths \cite{rf1975_Griffiths_multivMeix} provided their multivariate generalization, orthogonal with respect to the negative multinomial distribution. A group theoretical interpretation whereby Meixner polynomials in $d$ variables arise as matrix elements of $SO(d,1)$ representations on oscillator states was given recently \cite{rf2014_Genest_Miki_Vinet_Zhedanov_JPhysA_multivMeixInterp} and allowed for an elegant characterization of these polynomials. There remains to cast their $q$-analogs in a similar algebraic setting.

We thus wish to initiate for the $q$-Meixner polynomials a program that has been carried through in part for the $q$-Krawtchouk polynomials. Of the 4 families of $q$-Krawtchouk polynomials \cite{rf2010_Koekoek_Lesky_Swarttouw_hypergeometric}, two, the related quantum and affine $q$-Krawtchouk polynomials, have been given algebraic interpretations. At the univariate level, two dual points of view have been offered. On the one hand, Koornwinder obtained in \cite{rf1989_Koornwinder_repsuq2} the $q$-Krawtchouk polynomials as matrix elements of unitary $q$-irreducible representations of twisted $SU_q(2)$ quantum group. See also \cite{rf2000_Koelink_qkrawspherfct} and \cite{rf2004_Smirnov_Campigotto_qMeixAsDFctsu11}. A similar treatment of the $q$-Meixner polynomials based on the quantum group $SU_q(1,1)$ is given in \cite{rf2004_Smirnov_Campigotto_qMeixAsDFctsu11} in addition. On the other hand, Genest, Post, Vinet, Yu and Zhedanov \cite{rf2016_Genest_Post_Vinet_Yu_Zhedanov_qrotkraw} identified these same polynomials as matrix elements of $q$-exponentials of $\mathcal{U}_q(\mathfrak{su}(2))$ generators on representation bases. The latter approach was subsequently generalized \cite{rf2016_Genest_Post_Vinet_algebqkraw} to encompass the multivariate situation and to interpret the $q$-Krawtchouk polynomials in many variables introduced by Gasper and Rahman \cite{rf2007_Gasper_Rahman_multivqracah}.

We shall here focus on the univariate $q$-Meixner polynomials. This is interesting on its own and essential for a study of the multivariate case. Significant differences with respect to the analysis of the $q$-Krawtchouk polynomials given in \cite{rf2016_Genest_Post_Vinet_Yu_Zhedanov_qrotkraw} will be observed. An embedding of $\mathcal{U}_q(\mathfrak{su}(1,1))$ in the direct sum of two $q$-Weyl algebras associated to a pair of independent $q$-oscillators will be used and the $q$-Meixner polynomials will be observed to arise as matrix elements of unitary $q$-pseudorotation operators built from $q$-exponentials in the $\mathcal{U}_q(\mathfrak{su}(1,1))$ generators realized in terms of $q$-boson operators. While other treatments of the $q$-Meixner polynomials can be found elsewhere \cite{rf2004_Smirnov_Campigotto_qMeixAsDFctsu11, rf2003_Atakishiev_Atakishiev_Klimyk_qLaguerreqMeix, rf2000_Rosengren_quantalgebAskWil}, we wish to point out that the approach presented here extends itself naturally to the multivariable case. We shall also illustrate its power by using it to obtain a full characterization of the polynomials.

The article is organized as follows. In the first section, a model à la Schwinger of $\mathcal{U}_q(\mathfrak{su}(1,1))$ in terms of two independent $q$-oscillators is presented. In \autoref{sec:unitoper} the unitary operators of interest are introduced. Their matrix elements are then expressed in terms of $q$-Meixner polynomials in \autoref{sec:matelem}. Unitarity naturally leads to the orthogonality relations. In \autoref{sec:qto1limit}, the $q\to1$ limits of the unitary operators and of the matrix elements are given. The determination of the backward, forward relations and difference equation is carried out in \autoref{sec:backfordiff}. Complementary backward, forward relations and recurrence relation are presented in \autoref{sec:dualbackforrecu}. A duality property satisfied by the polynomials leading to new identities is exhibited in \autoref{sec:duality}. Generating functions of two types are derived in \autoref{sec:genfct}. Concluding remarks will be found in \autoref{sec:conclusion}. Basic $q$-identities that are used throughout the paper have been collected for reference in \autoref{app:AppendixA}. \autoref{app:AppendixB} contains the list of the ``dual'' relations satisfied by the $q$-Meixner polynomials.

\section{$\mathcal{U}_q(\mathfrak{su}(1,1))$ and $q$-oscillators}
\label{sec:model}
Consider two uncoupled $q$-oscillators \cite{rf1989_Macfarlane_qharmosc, rf1989_Biedenharn_quantgroupandqboson, rf1991_Floreanini_Vinet_qorthpolynoscillgroup} A and B whose sets of dynamical operators, $\{A\pm,A_0\}$ and $\{B\pm,B_0\}$, respectively, obey the following relations :\vspace{1em}

\begin{minipage}[t]{0.4\textwidth}
\center{\hspace{-3em}Oscillator A :}
\vspace{-1em}
\begin{align*}
\begin{aligned}
\vspace{0em}
[A_0 , A_\pm] &= \pm A_\pm~, \\
[A_- , A_+] &= q^{A_0}~, \\
A_-A_+ - q&A_+A_- = 1~,
\end{aligned}
\end{align*}
\end{minipage}
\begin{minipage}[t]{0.54\textwidth}
\center{\hspace{-8em}Oscillator B :} 
\vspace{-1em}
\begin{align}
\label{eqABalgebra}
\begin{aligned}
\vspace{0em}
[B_0 , B_\pm] &= \pm B_\pm~, \\
[B_- , B_+] &= q^{-B_0-1}~, \\
qB_-B_+ - &B_+B_- = 1~,
\end{aligned}
\end{align}
\end{minipage}
\begin{align}
\text{with}\quad[A_\pm,B_\pm]=[A_\pm,B_0]=[A_0,B_\pm]=[A_0,B_0]=0~.\nn
\end{align}
This can be viewed as a two-dimensional system. The subalgebras associated to the $q$-oscillators A and B admit the semi-infinite irreducible representations given by the following actions on the orthonormal basis vectors $\ket{n_A}$, $\ket{n_B}$ with $n_A$, $n_B=0,1,2,\dots$ :

\begin{minipage}[t]{0.4\textwidth}
\vspace{-2em}
\begin{align*}
\begin{aligned}
A_0\ket{n_A} &= n_A\ket{n_A}~,\\
A_-\ket{n_A} &= \sqrt{\frac{1-q^{n_A}}{1-q}} \ket{n_A-1}~, \\
A_+\ket{n_A} &= \sqrt{\frac{1-q^{n_A+1}}{1-q}} \ket{n_A+1}~, 
\end{aligned}
\end{align*}
\end{minipage}
\begin{minipage}[t]{0.54\textwidth}
\vspace{-2em}
\begin{align}
\label{eqABactions}
\begin{aligned}
B_0\ket{n_B} &= n_B\ket{n_B}~,\\
B_-\ket{n_B} &= \sqrt{\frac{q^{-n_B}-1}{1-q}} \ket{n_B-1}~, \\
B_+\ket{n_B} &= \sqrt{\frac{q^{-(n_B+1)}-1}{1-q}} \ket{n_B+1}~. 
\end{aligned}
\end{align}
\end{minipage}\\
Note that $A_\pm^\dagger = A_\mp$ and $B_\pm^\dagger = B_\mp$ in this representation.

Consider the operators
\begin{align}
\label{eqJoper}
J_0 = \frac{A_0 + B_0 + 1}{2} \qquad ,\qquad J_\pm = q^{\frac{B_0-A_0+2}{2}}A_\pm B_\pm~.
\end{align}
They act on the vector space spanned by the basis states $\ket{n_A} \otimes \ket{n_B}\equiv\ket{n_A,n_B}$ of the combined system. It is immediate to check that $J_0$, $J_\pm$ realize the commutation relations of $\mathcal{U}_q(\mathfrak{su}(1,1))$ :
\begin{align}
\label{eqJalgebra}
\begin{aligned}
\vspace{0em}
[J_0,J_\pm]&=\pm J_\pm~\qquad,\qquad[J_+,J_-]&=-\frac{q^{J_0}-q^{-J_0}}{q^{1/2}-q^{-1/2}}~.
\end{aligned}
\end{align}
It will prove practical to use the following notation for the 2 $q$-oscillator vector states. We shall write
\begin{align}
\ket{n}_\beta\equiv\ket{n,n+\beta-1}\quad,\quad\beta=1,2,3,\dots
\end{align}
where
\begin{align}
n\equiv n_A\qquad,\qquad\beta\equiv n_B-n_A+1~.
\end{align}
It is immediate to see that the orthonormality relation
\begin{align}
\label{eqOrthonV}
{}_{\beta}\brkt{n}{n^{\prime}}_{\beta^{\prime}}=\delta_{\beta\beta^{\prime}}\delta_{nn^{\prime}}
\end{align}
follows from $\brkt{n_A, n_B}{n_A^{\prime},n_B^{\prime}}=\delta_{n_A n'_A}\delta_{n_B n_B^{\prime}}$. It is also readily observed that $J_0$ and $J_\pm$ preserve the value of $\beta$. As a matter of fact, the actions of these operators in the basis $\ket{n}_\beta$ read :
\begin{align}
\label{eqJactions}
\begin{aligned}
J_0\ket{n}_\beta&=\left(n+\frac{\beta}{2}\right)\ket{n}_\beta~, \\
J_-\ket{n}_\beta&=q^{\frac{\beta+1}{2}}\sqrt{\frac{1-q^n}{1-q} \frac{q^{1-n-\beta}-1}{1-q}} \ket{n-1}_\beta~, \\
J_+\ket{n}_\beta&=q^{\frac{\beta+1}{2}}\sqrt{\frac{1-q^{n+1}}{1-q} \frac{q^{-n-\beta}-1}{1-q}} \ket{n+1}_\beta~.
\end{aligned}
\end{align}
One thus sees that $J_0$ and $J_\pm$ transform among themselves the vector states $\ket{n}_\beta$ with a fixed value of $\beta$. An infinite-dimensional representation of $\mathcal{U}_q(\mathfrak{su}(1,1))$ labelled by $\beta$ has thus been constructed.

\section{The unitary operator $U(\theta)$ and its matrix elements}
\label{sec:unitoper}
In this section, a $q$-analog of the unitary operator representing $SU(1,1)$ group elements is introduced and its unitarity is demonstrated. Its matrix elements will prove related to $q$-Meixner polynomials. Operators of that type have also been introduced by Zhedanov \cite{rf1993_Zhedanov_qrot}.

\subsection{The unitary operator $U(\theta)$}
Consider the following operator $U(\theta)$ :
\begin{align}
\label{eqOperU}
\hspace{-2em}U(\theta)=e_q^{1/2}\left(-\theta^2q^{-A_0}\right) e_q\left(\theta(1-q)q^{\frac{B_0-A_0+1}{2}}A_+B_+\right) E_q\left(-\theta(1-q)q^{\frac{B_0-A_0+1}{2}}A_-B_-\right) E_q^{1/2}\left(\theta^2q^{B_0+1}\right)~.
\end{align}
The limit of this operator when $q\to1$ will be obtained in \autoref{sec:qto1limit} and the result will justify the statement made at the beginning of this section.

Let us now show that $U(\theta)$ is unitary. In the framework of the previous sections, using Eqs. \eqref{eqABalgebra} and \eqref{eqBCH2}, one can obtain the following relations :
\begin{align}
e_q(\alpha A_-B_-)(A_+B_+)E_q(-\alpha A_-B_-) = A_+B_+-\frac{\alpha q^{A_0}}{(1-q)^2} + \frac{\alpha q^{-B_0-1}}{(1-q)^2(1-\alpha A_-B_-)}
\end{align}
and
\begin{align}
e_q(\alpha A_-B_-)q^{-B_0-1}E_q(-\alpha A_-B_-) = \frac{q^{-B_0-1}}{1-\alpha A_-B_-}~.
\end{align}
This leads to
\begin{align}
e_q(\alpha A_-B_-)\left(A_+B_+ -\frac{\alpha q^{-B_0-1}}{(1-q)^2}\right)E_q(-\alpha A_-B_-) = A_+B_+ -\frac{\alpha q^{A_0}}{(1-q)^2}~,
\end{align}
which allows one to write
\begin{align}
e_q(\alpha A_-B_-)e_q\left(\beta A_+B_+ -\frac{\alpha \beta q^{-B_0-1}}{(1-q)^2}\right) = e_q\left(\beta A_+B_+ -\frac{\alpha \beta q^{A_0}}{(1-q)^2}\right) e_q(\alpha A_-B_-)~.
\end{align}
Remarking that
\begin{align}
(A_+B_+)(q^{-B_0-1})&=q(q^{-B_0-1})(A_+B_+)~,\nl
(q^{A_0})(A_+B_+)&=q(A_+B_+)(q^{A_0})~,\nn
\end{align}
and using \eqref{eqeqsum}, one obtains the identity
\begin{align}
\label{eqIdeee}
e_q(\alpha A_-B_-)e_q\left(-\frac{\alpha \beta q^{-B_0-1}}{(1-q)^2}\right)e_q(\beta A_+B_+) = e_q(\beta A_+B_+)e_q\left(-\frac{\alpha \beta q^{A_0}}{(1-q)^2}\right)e_q(\alpha A_-B_-)~.
\end{align}
Similarly, inversing the previous equation, one finds
\begin{align}
\label{eqIdEEE}
E_q(\gamma A_+B_+)E_q\left(\frac{\gamma \delta q^{-B_0-1}}{(1-q)^2}\right)E_q(\delta A_-B_-) =E_q(\delta A_-B_-)E_q\left(\frac{\gamma \delta q^{A_0}}{(1-q)^2}\right)E_q(\gamma A_+B_+)~.
\end{align}

The fact that \eqref{eqOperU} is a unitary operator ($UU^\dagger=U^\dagger U=1$) can now easily be checked with the help of \eqref{eqIdeee} and \eqref{eqIdEEE}.

\section{Matrix elements of $U(\theta)$ and $q$-Meixner polynomials}
\label{sec:matelem}
We now wish to determine the matrix elements of the $q$-pseudorotation operator $U(\theta)$, in the basis formed by the $q$-oscillator states $\ket{x}_\beta$.

Using the fact that
\begin{align}
\label{eqABaction}
(A_-B_-)^\mu \ket{x}_\beta &= (1-q)^{-\mu} \sqrt{(q^{-x};q)_\mu\ (q^{1-x-\beta};q)_\mu q^{\mu x - \binom{\mu}{2}}}
\ket{x-\mu}_\beta~, \\
(A_+B_+)^\nu \ket{y}_\beta &= (1-q)^{-\nu} \sqrt{(q^{y+1};q)_\nu\ (q^{y+\beta};q)_\nu q^{-\nu(y+\beta)-\binom{\nu}{2}}}
\ket{y+\nu}_\beta~, 
\end{align}
and the definition \eqref{eqqExp}, one finds :
\begin{align}
\xi_{n,x}^{(\beta)}(\theta)&={}_{\beta}\bra{n}U(\theta)\ket{x}_{\beta}\nl
&=E_q^{1/2}\left(\theta^2q^{x+\beta}\right)\sum_{\mu}\sum_{\nu}e_q^{1/2}\left(-\theta^2q^{-(x-\mu+\nu)}\right)\frac{\theta^{\mu+\nu}(-1)^{\mu}}{(q;q)_\mu(q;q)_\nu}\sqrt{(q^{-x};q)_\mu(q^{1-x-\beta};q)_\mu}\nl
&\times\sqrt{(q^{x-\mu+1};q)_\nu(q^{x-\mu+\beta};q)_\nu}\sqrt{q^{\binom{\mu}{2}-\binom{\nu}{2}+\mu x+\mu\beta+\nu\mu-\nu x}}{}_{\beta}\brkt{n}{x-\mu+\nu}_{\beta}~.
\end{align}
In terms of the variable $\gamma=x-\mu$, the orthogonality of the basis vectors \eqref{eqOrthonV} imposes $n=x-\mu+\nu=\gamma+\nu$. Further simplifications using various $q$-Pochhammer identities lead to
\begin{align}
\label{eqXi}
\hspace{-2em}\xi_{n,x}^{(\beta)}=(-1)^x \theta^{n+x} &{n+\beta-1 \brack n}_q^{\frac12} {x+\beta-1 \brack x}_q^{\frac12} \sqrt{\frac{q^{\binom{x}{2}-\binom{n}{2}}}{(-\theta^2;q)_{x+\beta}(-\theta^2q^{-n};q)_n}} \nl
&\times\sum_{\gamma}\frac{\left(q^{-n};q\right)_\gamma\left(q^{-x};q\right)_\gamma}{(q;q)_\gamma\left(q^\beta;q\right)_\gamma}\left(-\frac{q^{n+1}}{\theta^2}\right)^\gamma~.
\end{align}
Recalling definition \eqref{eqHyperMeix}, one arrives at
\begin{align}
\label{eqXii}
\hspace{-2em}\xi_{n,x}^{(\beta)}=(-1)^x \theta^{n+x} {n+\beta-1 \brack n}_q^{\frac12} {x+\beta-1 \brack x}_q^{\frac12} \sqrt{\frac{q^{\binom{x}{2}-\binom{n}{2}}}{(-\theta^2;q)_{x+\beta}(-\theta^2q^{-n};q)_n}} \mathcal{M}_n\left(q^{-x};q^{\beta-1},\theta^2;q\right)~,
\end{align}
which entails the interpretation of $q$-Meixner polynomials as matrix elements of $q$-pseudorotation representations on $q$-oscillator states (see also \cite{rf1993_Zhedanov_qrot}). Note that these matrix elements are real and that in general the formulas derived within the present setting $($with $b=q^{\beta-1})$ extend to the full admissible range of the parameter $b$ $(0<b<1)$ as in \cite{rf2010_Koekoek_Lesky_Swarttouw_hypergeometric}.

\subsection{Orthogonality relations}
The unitarity of the operator $U(\theta)$ can be used to obtain the orthogonality of the $q$-Meixner polynomials in the following way.

Introduce the negative binomial distribution
\begin{align}
\label{eqOmega}
\omega_x^{(\beta)}=\left[\xi_{0,x}^{(\beta)}\right]^2=\theta^{2x}{x+\beta-1 \brack x}_q\frac{q^{\binom{x}{2}}}{(-\theta^2;q)_{x+\beta}}~.
\end{align}
One can write
\begin{align}
\label{eqOrthMatx}
{_\beta\bra{n'}}U U^\dagger\ket{n}_\beta = \sum_{x=0}^\infty {_\beta\bra{n'}}U\ket{x}_\beta {_\beta\bra{x}}U^\dagger\ket{n}_\beta 
= \sum_{x=0}^\infty \xi_{n',x}^{(\beta)} \left(\xi_{n,x}^{(\beta)}\right)^* = \delta_{nn^\prime}~.
\end{align}
Substituting \eqref{eqXii} and using the reality of the matrix elements, the relation \eqref{eqOrthMatx} translates into the orthogonality relation for the $q$-Meixner polynomials :
\begin{align}  
\label{eqOrthPolyn}
\hspace{-2em}\sum_{x=0}^\infty\omega_x^{(\beta)}\mathcal{M}_n\left(q^{-x};q^{\beta-1},\theta^2;q\right)\mathcal{M}_{n^\prime}\left(q^{-x};q^{\beta-1},\theta^2;q\right)=\frac{q^{\binom{n}{2}}}{\theta^{2n}}\frac{(q;q)_n\left(-\theta^2 q^{-n};q\right)_n}{(q;q)_{n+\beta-1}(q;q)_{\beta-1}}\delta_{nn^\prime}~.
\end{align}
A dual relation can similarly be obtained. One has
\begin{align}
\label{eqOrthMatn}
{_\beta\bra{x'}}U^\dagger U\ket{x}_\beta = \sum_{n=0}^\infty {_\beta\bra{x'}}U^\dagger\ket{n}_\beta {_\beta\bra{n}}U\ket{x}_\beta 
= \sum_{n=0}^\infty \xi_{n,x}^{(\beta)} \left(\xi_{n,x'}^{(\beta)}\right)^* = \delta_{xx^\prime}~.
\end{align}
Following parallel steps, the following relation is found :
\begin{align}
\label{eqOrthPolynDual}
\hspace{-2em}\sum_{n=0}^\infty\mathcal{M}_n\left(q^{-x};q^{\beta-1},\theta^2;q\right)\mathcal{M}_n\left(q^{-x^\prime};q^{\beta-1},\theta^2;q\right)\frac{\theta^{2n}q^{-\binom{n}{2}}}{\left(-\theta^2 q^{-n};q\right)_n}{n+\beta-1 \brack n}_q=\frac{1}{\omega_x^{(\beta)}}\delta_{xx^\prime}~.
\end{align}

\section{$q\to1$ limit} 
\label{sec:qto1limit}
It is instructive to obtain the $q\to1$ limits of $U(\theta)$ and its matrix elements. This is done now.

\subsection{Operator $U(\theta)$}
\label{subsec:operU}
Let
\begin{align}
\lim_{q\to 1}X_0 = \widetilde{X_0}~,\qquad \lim_{q\to 1}X_\pm = \widetilde{X_\pm}~, \qquad  X = A,B~.
\end{align}
In the limit $q\to1$, the commutation relations become :
\begin{align}
\begin{aligned}
\label{eqqto1algebra}
&[\widetilde{X_0},\widetilde{X_\pm}] = \pm\widetilde{X_\pm}~,\qquad [\widetilde{X_-},\widetilde{X_+}] = 1~,\qquad X = A,B~,\\ 
&[\widetilde{A}_\pm,\widetilde{B}_\pm]=[\widetilde{A}_\pm,\widetilde{B}_0]=[\widetilde{A}_0,\widetilde{B}_\pm]=[\widetilde{A}_0,\widetilde{B}_0]=0~.
\end{aligned}
\end{align}
Noting that
\begin{align}
&\lim_{q\to 1}\frac{e_q\left(-\theta^2 q^{-B_0-1} \right)}{e_q\left(-\theta^2 \right)} = \left(1+\theta^2\right)^{-(1+\widetilde{B_0})} 
= \exp\left(-\ln(1+\theta^2)(1+\widetilde{B_0})\right)~, \\
&\lim_{q\to 1}\frac{E_q\left(\theta^2 q^{A_0} \right)}{E_q\left(\theta^2 \right)} = \left(1+\theta^2\right)^{-\widetilde{A_0}} 
= \exp\left(-\ln(1+\theta^2)\widetilde{A_0}\right)~,
\end{align}
the $q\to 1$ limit of \eqref{eqOperU}, denoted $\widetilde{U(\theta)}$, is found to be :
\begin{align}
\label{eqLimOp1}
\hspace{-2em}\widetilde{U(\theta)}= \exp\left[-\ln(1+\theta^2)\frac{(1+\widetilde{B_0})}{2}\right] \exp\left[\theta\widetilde{A_+}\widetilde{B_+}\right]\exp\left[-\theta\widetilde{A_-}\widetilde{B_-}\right] \exp\left[-\ln(1+\theta^2)\frac{\widetilde{A_0}}{2}\right]~.
\end{align}
The identities
\begin{align}
\label{eqIdBCHcl}
e^{\alpha \widetilde{A_-}}e^{-\beta \widetilde{A_0}}=e^{-\beta \widetilde{A_0}}e^{(e^{-\beta})\alpha\widetilde{A_-}}~,\qquad \text{and} \qquad e^{\gamma \widetilde{B_0}}e^{\delta \widetilde{B_+}}=e^{(e^{\gamma})\delta \widetilde{B_+}}e^{\gamma\widetilde{B_0}}~,
\end{align}
obtained from the usual Baker–Campbell–Hausdorff formula $e^A B e^{-A} = B + [A,B] + \frac12[A,[A,B]] + \cdots$ allow to simplify \eqref{eqLimOp1} into
\begin{align}
\label{eqLimOp2}
\hspace{-1em}\widetilde{U(\theta)} = \exp\left[\frac{\theta}{\sqrt{1+\theta^2}}\widetilde{A_+}\widetilde{B_+}\right]\exp\left[-\ln(1+\theta^2)\frac{(\widetilde{A_0}+\widetilde{B_0}+1)}{2}\right]\exp\left[\frac{-\theta}{\sqrt{1+\theta^2}}\widetilde{A_-}\widetilde{B_-}\right]~.
\end{align}
Recall the Schwinger realization of $\mathfrak{su}(1,1)$ :
\begin{align}
\label{eqClassSu11}
\widetilde{J_0} = \frac{\widetilde{A_0}+\widetilde{B_0}+1}{2} \quad &, \quad \widetilde{J_\pm} = \widetilde{A_\pm}\widetilde{B_\pm} \quad ,\\
[\widetilde{J_0},\widetilde{J_\pm}] = \pm \widetilde{J}_\pm \quad&,\quad [\widetilde{J_+},\widetilde{J_-}]=-2\widetilde{J_0}~. \nn
\end{align}
Under the change of variable $\theta = \sinh \tau$, $\tau \in \mathbb{R}$, \eqref{eqLimOp2} becomes
\begin{align}
\label{eqUClassInterp}
\widetilde{U}(\sinh \tau) &= \exp\left[\tanh \tau \widetilde{J_+}\right] \exp\left[-2 \ln(\cosh\tau)\widetilde{J_0}\right] \exp\left[-\tanh \tau \widetilde{J_-}\right] \nl
&= \exp\left[\tau\left(\widetilde{J_+} - \widetilde{J_-}\right)\right]~,
\end{align}
where the following disentangling formula for $\mathfrak{su}(1,1)$ \cite{rf1985_Truax_squeeze} has been used :
\begin{align}
\label{eqsu11Truax}
\hspace{-2em}\exp\left(\tau L_+-\overline{\tau}L_-\right)=\exp\left[\left(\frac{\tau}{|\tau|}\tanh|\tau|\right)L_+\right]\exp\left[-2\ln\left(\cosh|\tau|\right)L_0\right]\exp\left[-\left(\frac{\overline{\tau}}{|\tau|}\tanh|\tau|\right)L_-\right]~.
\end{align}

The operator \eqref{eqUClassInterp} is hence identified as representing a (two-dimensional) pseudorotation, thus allowing to say that $U(\theta)$ represents a $q$-pseudorotation in two dimensions.

\subsection{Matrix elements}
\label{subsec:matrixelem}
For the $q$-Meixner polynomials one has \cite{rf2010_Koekoek_Lesky_Swarttouw_hypergeometric}
\begin{align}
\label{eqLimQMeixMeix}
\lim_{q\to 1}\mathcal{M}_n\left(q^{-x};q^{\beta-1},\frac{c}{1-c};q\right)=M_n(x;\beta,c)~.
\end{align}
One thus straightforwardly obtains :
\begin{align}
\label{eqLimXi1}
\lim_{q\to 1}\xi_{n,x}^{(\beta)}= (-1)^x (\theta)^{x+n}{n+\beta-1 \choose n}^{\frac12} {x+\beta-1 \choose x}^{\frac12} \left(1+\theta^2\right)^{-\frac{(\beta+n+x)}{2}} M_n\left(x,\beta,\frac{\theta^2}{1+\theta^2}\right)~.
\end{align}
Under the change of variable $\theta = \sinh \tau$, this becomes
\begin{align}
\label{eqLimXi2}
\lim_{q\to 1}\xi_{n,x}^{(\beta)}= (-1)^x {n+\beta-1 \choose n}^{\frac12} {x+\beta-1 \choose x}^{\frac12} \frac{(\tanh\tau)^{x+n}}{(\cosh\tau)^\beta} M_n\left(x,\beta,\tanh^2\tau\right)~,
\end{align}
in keeping with results obtained in \cite{rf2014_Genest_Miki_Vinet_Zhedanov_JPhysA_multivMeixInterp} on the $SU(1,1)$ interpretation of the standard univariate Meixner polynomials.

\section{Backward, forward relations and difference relation}
\label{sec:backfordiff}
The algebraic interpretation of the $q$-Meixner polynomials that we have provided offers a cogent framework to derive the basic features of those polynomials. We shall focus in this section on the lowering and raising formulas as well as the difference equation. We shall begin by deriving two identities that will prove fundamental in obtaining the desired properties. We wish to show that
\begin{align}
\label{eqUdAmU}
U^\dagger(q^{-1/2}\theta)A_-U(\theta)=A_-\sqrt{1+\theta^2 q^{B_0}}+\theta q^{\frac{A_0+B_0}{2}}B_+~.
\end{align}
First, from the commutation relations \eqref{eqABalgebra}, one finds that
\begin{align}
\hspace{-2em}U^\dagger(q^{-1/2}\theta)A_-U(&\theta)=E_q^{1/2}\left(\theta^2q^{B_0}\right)E_q\left(-\theta(1-q)q^{\frac{B_0-A_0}{2}}A_+B_+\right)A_-\nl
&\times\Bigg[e_q\left(\theta(1-q)q^{\frac{B_0-A_0+1}{2}}A_-B_-\right)e_q\left(-\theta^2q^{-A_0}\right)e_q\left(\theta(1-q)q^{\frac{B_0-A_0+1}{2}}A_+B_+\right)\Bigg]\nl
&\times E_q\left(-\theta(1-q)q^{\frac{B_0-A_0+1}{2}}A_-B_-\right) E_q^{1/2}\left(\theta^2q^{B_0+1}\right)~.
\end{align}
Using \eqref{eqIdeee}, this leads to
\begin{align}
U^\dagger(q^{-1/2}\theta)&A_-U(\theta)=E_q^{1/2}\left(\theta^2q^{B_0}\right)E_q\left(-\theta(1-q)q^{\frac{B_0-A_0}{2}}A_+B_+\right)A_-\nl
&\hspace{-2em}\times\Bigg[e_q\left(\theta(1-q)q^{\frac{B_0-A_0+1}{2}}A_+B_+\right)e_q\left(-\theta^2q^{B_0+1}\right)e_q\left(\theta(1-q)q^{\frac{B_0-A_0+1}{2}}A_-B_-\right)\Bigg]\nl
&\hspace{-2em}\times E_q\left(-\theta(1-q)q^{\frac{B_0-A_0+1}{2}}A_-B_-\right)E_q^{1/2}\left(\theta^2q^{B_0+1}\right)\nl
&\hspace{-2em}=E_q^{1/2}\left(\theta^2q^{B_0}\right) \Bigg[E_q\left(-\theta(1-q)q^{\frac{B_0-A_0}{2}}A_+B_+\right)A_- e_q\left(\theta(1-q)q^{\frac{B_0-A_0+1}{2}}A_+B_+\right)\Bigg]\nl
&\hspace{-2em}\times e_q^{1/2}\left(-\theta^2q^{B_0+1}\right)~.\nn
\end{align}
With the help of \eqref{eqBCH1}, the expression in the square brackets is computed to be
\begin{align}
E_q\left(-\theta(1-q)q^{\frac{B_0-A_0}{2}}A_+B_+\right)A_- e_q\left(\theta(1-q)q^{\frac{B_0-A_0+1}{2}}A_+B_+\right) = A_-+\theta q^{\frac{A_0+B_0}{2}}B_+~,
\end{align}
and one thus arrives at \eqref{eqUdAmU}.

The Hermitian conjugate of the relation \eqref{eqUdAmU} also gives another useful identity :
\begin{align}
\label{eqUdApU}
U^\dagger(\theta)A_+U(q^{-1/2}\theta)=A_+\sqrt{1+\theta^2 q^{B_0}}+\theta B_- q^{\frac{A_0+B_0}{2}}~.
\end{align}

\subsection{Backward relation}
The backward relation is obtained as follows. One has
\begin{align}
{}_{\beta+1}\bra{n}A_-U(\theta)\ket{x}_{\beta}&=\sqrt{\frac{1-q^{n+1}}{1-q}}\xi_{n+1,x}^{(\beta)}(\theta)\nl
&={}_{\beta+1}\bra{n}U(q^{-1/2}\theta)U^\dagger(q^{-1/2}\theta)A_-U(\theta)\ket{x}_{\beta}\nl
&={}_{\beta+1}\bra{n}U(q^{-1/2}\theta)\Bigg(A_-\sqrt{1+\theta^2 q^{B_0}}+\theta q^{\frac{A_0+B_0}{2}}B_+\Bigg)\ket{x}_{\beta}~,
\end{align}
with the help of \eqref{eqUdAmU}. In terms of matrix elements, one finds
\begin{align}
\hspace{-2em}\sqrt{1-q^{n+1}}\xi_{n+1,x}^{(\beta)}(\theta)=\sqrt{\left(1-q^x\right)\left(1+\theta^2q^{x+\beta-1}\right)}\xi_{n,x-1}^{(\beta+1)}(q^{\nicefrac{-1}{2}}\theta)+\theta q^{\nicefrac{x}{2}}\sqrt{1-q^{x+\beta}}\xi_{n,x}^{(\beta+1)}(q^{\nicefrac{-1}{2}}\theta)~.
\end{align}
Using the expression \eqref{eqXii} of $\xi_{n,x}^{(\beta)}$ in terms of $q$-Meixner polynomials and simplifying, one arrives at
\begin{align}
\label{eqBackRel}
\hspace{-1em}\theta^2\left(1-q^\beta\right)\mathcal{M}_{n+1}\left(q^{-x};q^{\beta-1},\theta^2;q\right)&= q\left(1-q^{-x}\right) \left(1+\theta^2q^{x+\beta-1}\right)\mathcal{M}_{n}\left(q^{-(x-1)};q^{\beta},\theta^2q^{-1};q\right)\nl
&+\theta^2\left(1-q^{x+\beta}\right)\mathcal{M}_{n}\left(q^{-x};q^{\beta},\theta^2q^{-1};q\right)~,
\end{align}
which coincides with the relation given in \cite{rf2010_Koekoek_Lesky_Swarttouw_hypergeometric}.

\subsection{Forward relation}
One proceeds similarly for the forward relation. Using \eqref{eqUdApU}, we have
\begin{align}
{}_{\beta}\bra{n}A_+U(q^{-1/2}\theta)\ket{x}_{\beta+1}&=\sqrt{\frac{1-q^n}{1-q}}\xi_{n-1,x}^{(\beta+1)}(q^{-1/2}\theta)\nl
&={}_{\beta}\bra{n}U(\theta)\Bigg(A_+\sqrt{1+\theta^2 q^{B_0}}+\theta B_- q^{\frac{A_0+B_0}{2}}\Bigg)\ket{x}_{\beta+1}~.
\end{align}
Applying the $q$-oscillator operators on the right leads to
\begin{align}
\sqrt{1-q^n}\xi_{n-1,x}^{(\beta+1)}(q^{-1/2}\theta)=\sqrt{1+\theta^2q^{x+\beta}}\sqrt{1-q^{x+1}}\xi_{n,x+1}^{(\beta)}(\theta)+\theta q^{x/2}\sqrt{1-q^{x+\beta}}\xi_{n,x}^{(\beta)}(\theta)~.
\end{align}
Using \eqref{eqXii} and simplifying, the following forward relation for the $q$-Meixner polynomials is obtained :
\begin{align}
\label{eqForRel}
\hspace{-1em}\frac{1}{\theta^2q^x}\frac{1-q^n}{1-q^\beta}\mathcal{M}_{n-1}\left(q^{-x};q^\beta,\theta^2q^{-1};q\right)=\mathcal{M}_n\left(q^{-x};q^{\beta-1},\theta^2;q\right)-\mathcal{M}_n\left(q^{-(x+1)};q^{\beta-1};\theta^2;q\right)~,
\end{align}
which checks with the formulas in \cite{rf2010_Koekoek_Lesky_Swarttouw_hypergeometric}.

\subsection{Difference equation}
The difference equation is found by combining the two ladder relations. Indeed, we see that
\begin{align}
&{}_{\beta}\bra{n}A_+A_-U(\theta)\ket{x}_{\beta}=\frac{1-q^n}{1-q}\xi_{n,x}^{(\beta)}(\theta)\nl
&={}_{\beta}\bra{n}U(\theta)\Bigg(A_+\sqrt{1+\theta^2 q^{B_0}}+\theta B_- q^{\frac{A_0+B_0}{2}}\Bigg) \Bigg(A_-\sqrt{1+\theta^2 q^{B_0}}+\theta q^{\frac{A_0+B_0}{2}}B_+\Bigg)\ket{x}_{\beta}~,
\end{align}
upon using the identities \eqref{eqUdAmU} and \eqref{eqUdApU}. This leads to the following relation in terms of matrix elements
\begin{align}
\left(1-q^n\right)\xi_{n,x}^{(\beta)}(\theta)&=\left[ \left(1-q^x\right)\left(1+\theta^2q^{x+\beta-1}\right)+\theta^2q^x\left(1-q^{x+\beta}\right) \right]\xi_{n,x}^{(\beta)}(\theta)\nl
&+\theta\sqrt{1-q^{x+1}}\sqrt{1+\theta^2q^{x+\beta}}\sqrt{1-q^{x+\beta}}q^{x/2}\xi_{n,x+1}^{(\beta)}(\theta)\nl
&+\theta\sqrt{1-q^x}\sqrt{1+\theta^2q^{x+\beta-1}}\sqrt{1-q^{x+\beta-1}}q^{(x-1)/2}\xi_{n,x-1}^{(\beta)}(\theta)~.
\end{align}
Finally, given \eqref{eqXii}, the difference equation of the $q$-Meixner polynomials is obtained, and is seen to correspond to the one in \cite{rf2010_Koekoek_Lesky_Swarttouw_hypergeometric}
\begin{align}
\label{eqDiffRel}
\hspace{-2em}\left(1-q^n\right)\mathcal{M}_n\left(q^{-x};q^{\beta-1},\theta^2;q\right)&=-\theta^2q^x\left(1-q^{x+\beta}\right)\mathcal{M}_n\left(q^{-(x+1)};q^{\beta-1},\theta^2;q\right)\nl
&\hspace{-2em}+\left[ \left(1-q^x\right)\left(1+\theta^2q^{x+\beta-1}\right)+\theta^2q^x\left(1-q^{x+\beta}\right) \right]\mathcal{M}_n\left(q^{-x};q^{\beta-1},\theta^2;q\right)\nl
&\hspace{-2em}-\left(1-q^x\right)\left(1+\theta^2q^{x+\beta-1}\right)\mathcal{M}_n\left(q^{-(x-1)};q^{\beta-1},\theta^2;q\right)~.
\end{align}

\section{Complementary backward, forward relations and recurrence relation}
\label{sec:dualbackforrecu}
The recurrence relation and complementary ladder relations on the variable are derived by following an approach similar to the one of the last section. In this case, one needs formulas analogous to \eqref{eqUdAmU} and \eqref{eqUdApU} with $U(\theta)$ instead of $U^\dagger(\theta)$ on the left. With the help of \eqref{eqIdEEE}, one finds
\begin{align}
U(\theta)&q^{-A_0/2}A_-U^\dagger(\theta)=e_q^{1/2}\left(-\theta^2q^{-A_0}\right) e_q\left(\theta(1-q)q^{\frac{B_0-A_0+1}{2}}A_+B_+\right)q^{-A_0/2}A_- \nl
&\times\left[E_q\left(-\theta(1-q)q^{\frac{B_0-A_0+1}{2}}A_-B_-\right) E_q\left(\theta^2q^{B_0+1}\right)E_q\left(-\theta(1-q)q^{\frac{B_0-A_0+1}{2}}A_+B_+\right)\right]\nl
&\times e_q\left(\theta(1-q)q^{\frac{B_0-A_0+1}{2}}A_-B_-\right)e_q^{1/2}\left(-\theta^2q^{-A_0}\right)\nl
&=e_q^{1/2}\left(-\theta^2q^{-A_0}\right) e_q\left(\theta(1-q)q^{\frac{B_0-A_0+1}{2}}A_+B_+\right)q^{-A_0/2}A_- \nl
&\times\left[E_q\left(-\theta(1-q)q^{\frac{B_0-A_0+1}{2}}A_+B_+\right) E_q\left(\theta^2q^{-A_0}\right)E_q\left(-\theta(1-q)q^{\frac{B_0-A_0+1}{2}}A_-B_-\right)\right]\nl
&\times e_q\left(\theta(1-q)q^{\frac{B_0-A_0+1}{2}}A_-B_-\right)e_q^{1/2}\left(-\theta^2q^{-A_0}\right)\nl
&\hspace{-2em}=e_q^{1/2}\left(-\theta^2q^{-A_0}\right)\left\{e_q\left(\theta(1-q)q^{\frac{B_0-A_0+1}{2}}A_+B_+\right)q^{-A_0/2}A_-E_q\left(-\theta(1-q)q^{\frac{B_0-A_0+1}{2}}A_+B_+\right)\right\}\nl
&\hspace{-1em}\times E_q^{1/2}\left(\theta^2q^{-A_0}\right)~.\nn
\end{align}
Using the $q$-BCH formula \eqref{eqBCH2} and simplifying, the middle term in brackets is found to be equal to
\begin{align}
e_q\left(\theta(1-q)q^{\frac{B_0-A_0+1}{2}}A_+B_+\right)&q^{-A_0/2}A_-E_q\left(-\theta(1-q)q^{\frac{B_0-A_0+1}{2}}A_+B_+\right)\nl
&=q^{-A_0/2}A_--\theta q^{-A_0}q^{B_0/2}B_+~.
\end{align}
One thus arrives at the following identity
\begin{align}
\label{eqUAmUd}
U(\theta)q^{-A_0/2}A_-U^\dagger(\theta)=q^{-A_0/2}A_-\sqrt{1+\theta^2 q^{-A_0}}-\theta q^{-A_0}q^{B_0/2}B_+~.
\end{align}
The Hermitian conjugate of \eqref{eqUAmUd} gives a second one, which is also useful :
\begin{align}
\label{eqUApUd}
U(\theta)A_+q^{-A_0/2}U^\dagger(\theta)=\sqrt{1+\theta^2 q^{-A_0}}A_+q^{-A_0/2}-\theta q^{-A_0}B_-q^{B_0/2}~.
\end{align}

\subsection{Complementary backward relation}
A complementary backward relation is now derived as follows. It is seen that
\begin{align}
\hspace{-1em}\sqrt{1-q}\ {}_{\beta}\bra{n}U(\theta)q^{-A_0/2}A_-\ket{x}_{\beta-1}&=\sqrt{1-q^{x}}q^{-(x-1)/2}\xi_{n,x-1}^{(\beta)}(\theta)\nl
&\hspace{-12em}=\sqrt{1-q^{n+1}}\sqrt{1+\theta^2q^{-(n+1)}}q^{-n/2}\xi_{n+1,x}^{(\beta-1)}(\theta)-\theta\sqrt{1-q^{n+\beta-1}}q^{-n}\xi_{n,x}^{(\beta-1)}(\theta)~,
\end{align}
using \eqref{eqUAmUd}. Applying \eqref{eqXii}, one obtains the desired result
\begin{align}
\label{eqDualBack}
\hspace{-1em}\frac{q^{n+1}}{\theta^2}\frac{1-q^{-x}}{1-q^{\beta-1}}\mathcal{M}_{n}\left(q^{-(x-1)};q^{\beta-1},\theta^2;q\right)=\mathcal{M}_{n+1}\left(q^{-x};q^{\beta-2},\theta^2;q\right)-\mathcal{M}_{n}\left(q^{-x};q^{\beta-2},\theta^2;q\right)~.
\end{align}

\subsection{Complementary forward relation}
Here one starts from
\begin{align}
\sqrt{1-q}\ {}_{\beta}\bra{n}U(\theta)A_+q^{-A_0/2}\ket{x}_{\beta+1}&=\sqrt{1-q^{x+1}}q^{-x/2}\xi_{n,x+1}^{(\beta)}(\theta)\nl
&\hspace{-12em}=\sqrt{1-q^n}\sqrt{1+\theta^2q^{-n}}q^{-(n-1)/2}\xi_{n-1,x}^{(\beta+1)}(\theta)-\theta\sqrt{1-q^{n+\beta}}q^{-n}\xi_{n,x}^{(\beta+1)}(\theta)~,
\end{align}
where \eqref{eqUApUd} has been used. Applying \eqref{eqXii}, one then obtains
\begin{align}
\label{eqDualFor}
\theta^2q^n\left(1-q^\beta\right) \mathcal{M}_n\left(q^{-(x+1)};q^{\beta-1},\theta^2;q\right)&=\theta^2\left(1-q^{n+\beta}\right)\mathcal{M}_n\left(q^{-x};q^{\beta},\theta^2;q\right)\nl 
&-\left(q^n+\theta^2\right)\left(1-q^n\right)\mathcal{M}_{n-1}\left(q^{-x};q^{\beta},\theta^2;q\right)~.
\end{align}

\subsection{Recurrence equation}
The recurrence equation is found by combining the two previous relations. Note that
\begin{align}
&{}_{\beta}\bra{n}U(\theta)A_+ q^{-A_0}A_-\ket{x}_{\beta} = \frac{1-q^x}{1-q}q^{-(x-1)} \xi_{n,x}^{(\beta)}\nl
&={}_{\beta}\bra{n}\left(\sqrt{1+\theta^2 q^{-A_0}}A_+q^{-A_0/2}-\theta q^{-A_0}B_-q^{B_0/2}\right)\nl
&\hspace{13em}\times\left(q^{-A_0/2}A_-\sqrt{1+\theta^2 q^{-A_0}}-\theta q^{-A_0}q^{B_0/2}B_+\right)U(\theta)\ket{x}_{\beta}~. \nn
\end{align}
This leads to the following relation for the matrix elements
\begin{align}
\left(1-q^x\right)q^{-(x-1)}\xi_{n,x}^{(\beta)}&=\left[\left(1-q^n\right)\left(1+\theta^2q^{-n}\right)q^{-n+1} + \theta^2 q^{-2n}\left(1-q^{n+\beta}\right)\right] \xi_{n,x}^{(\beta)}\nl
&-\theta q^{-3n/2} \sqrt{1-q^{n+1}}\sqrt{1+\theta^2q^{-n-1}}\sqrt{1-q^{n+\beta}} \xi_{n+1,x}^{(\beta)}\nl
&-\theta q^{-3(n-1)/2} \sqrt{1-q^n}\sqrt{1+\theta^2q^{-n}}\sqrt{1-q^{n+\beta-1}} \xi_{n-1,x}^{(\beta)}~,
\end{align}
and calling upon \eqref{eqXii}, the recurrence relation \cite{rf2010_Koekoek_Lesky_Swarttouw_hypergeometric} is finally obtained :
\begin{align}
\label{eqReccRel}
\hspace{-2em}q^{2n+1} \left(1-q^{-x}\right)\mathcal{M}_n\left(q^{-x};q^{\beta-1},\theta^2;q\right) &= q\left(1-q^n\right)\left(q^n+\theta^2\right) \mathcal{M}_{n-1}\left(q^{-x};q^{\beta-1},\theta^2;q\right) \nl
&\hspace{-3em}-\left[q\left(1-q^n\right)\left(q^n+\theta^2\right) + \theta^2\left(1-q^{n+\beta}\right)\right] \mathcal{M}_n\left(q^{-x};q^{\beta-1},\theta^2;q\right) \nl
&\hspace{-3em}+\theta^2\left(1-q^{n+\beta}\right) \mathcal{M}_{n+1}\left(q^{-x};q^{\beta-1},\theta^2;q\right)~.
\end{align}

\section{Duality}
\label{sec:duality}
The following relation expresses the property of the $q$-Meixner polynomials under an exchange of the degree and the variable :
\begin{align}
\label{eqDualMeix}
\hspace{-2em}\mathcal{M}_n\left(q^{-x};q^{\beta-1},\theta^2;q\right)&={}_{2} \phi_1\left({{q^{-n},q^{-x}} \atop {q^\beta}} \middle| q;-\frac{q^{n+1}}{\theta^2} \right)\nl
&={}_{2} \phi_1\left({{q^{-x},q^{-n}} \atop {q^\beta}} \middle| q;-\frac{q^{x+1}}{\theta^2 q^{x-n}} \right)=\mathcal{M}_x\left(q^{-n};q^{\beta-1},\theta^2 q^{x-n};q\right)~.
\end{align}
In terms of matrix elements, this amounts to
\begin{align}
\label{eqDualXi}
\xi_{n,x}^{(\beta)}(\theta)=\sqrt{q^{n-x}}\ \xi_{x,n}^{(\beta)}(-\theta q^{(x-n)/2})~.
\end{align}

Using this duality property, other relations can be derived in a simple way. For example, starting from the recurrence relation \eqref{eqReccRel} and applying the above relation, one obtains
\begin{align}
\hspace{-2em}q^{2n+1}\left(1-q^{-x}\right)\mathcal{M}_{n}\left(q^{-x};q^{\beta-1},\theta^2q^{x-n};q\right)&=\theta^2\left(1-q^{n+\beta}\right)\mathcal{M}_{n+1}\left(q^{-x},q^{\beta-1},\theta^2q^{x-(n+1)};q\right)\nl
&\hspace{-7em}-\left[q\left(1-q^{n}\right)\left(q^{n}+\theta^2\right)+\theta^2\left(1-q^{n+\beta}\right)\right]\mathcal{M}_{n}\left(q^{-x};q^{\beta-1},\theta^2q^{x-n};q\right)\nl
&\hspace{-7em}+q\left(1-q^{n}\right)\left(q^{n}+\theta^2\right)\mathcal{M}_{n-1}\left(q^{-x},q^{\beta-1},\theta^2q^{x-(n-1)};q\right)~.
\end{align}
Exchanging $x\leftrightarrow n$ and then taking $\theta^2\to\theta^2q^{x-n}$, one gets 
\begin{align}
\hspace{-2em}q^{x+1}\left(1-q^{n}\right)\mathcal{M}_n\left(q^{-x};q^{\beta-1},\theta^2;q\right)&=-q\left(1-q^x\right)\left(q^n+\theta^2\right)\mathcal{M}_n\left(q^{-(x-1)};q^{\beta-1},\theta^2q;q\right)\nl
&\hspace{-6em}+\left[q\left(1-q^x\right)\left(q^n+\theta^2\right)+\theta^2\left(1-q^{x+\beta}\right) \right]\mathcal{M}_n\left(q^{-x};q^{\beta-1},\theta^2;q\right)\nl
&\hspace{-6em}-\theta^2\left(1-q^{x+\beta}\right)\mathcal{M}_n\left(q^{-(x+1)};q^{\beta-1},\theta^2q^{-1};q\right)~.
\end{align}
Note that this relation is not exactly the difference equation \eqref{eqDiffRel} because the parameters $\theta^2$ are affected by a factor $q$ in some of the polynomials.

The same process can be applied to the other relations that were derived to obtain new relations, but again these relations will have their parameters $\theta^2$ modified by some factors of $q$, which means that those relations will differ from the usual ones. A list of these ``dual'' relations is included in \autoref{app:AppendixA}.

\section{Generating functions}
\label{sec:genfct}
Two generating functions are now obtained from the algebraic picture : one with respect to the degrees and the other with respect to the variables.

\subsection{Generating function with respect to the degrees}
A useful identity that is proved from \eqref{eqIdeee} is the following
\begin{align}
\label{eqExp24}
E_q(\gamma A_-B_-)e_q(\delta A_+B_+)=e_q\left(\frac{\gamma\delta q^{-B_0-1}}{(1-q)^2}\right)e_q(\delta A_+B_+)E_q(\gamma A_-B_-)E_q\left(\frac{-\gamma\delta q^{A_0}}{(1-q)^2}\right)~.
\end{align}
Now introduce the operator
\begin{align}
V(\theta,t)=E_q\left(t(1-q)q^{\frac{B_0-A_0+1}{2}}A_-B_-\right)E_q^{1/2}\left(\theta^2q^{-A_0}\right)
\end{align}
and let $\mathcal{F}_1={}_{\beta}\bra{n}V(\theta,t)U(\theta)\ket{0}_{\beta}$. The generating function will be arrived at by obtaining two expressions for $\mathcal{F}_1$. Acting first with $V(\theta,t)$ on the left leads to
\begin{align}
\mathcal{F}_1&={}_{\beta}\bra{0}E_q\left(t(1-q)q^{\frac{B_0-A_0+1}{2}}A_-B_-\right)E_q^{1/2}\left(\theta^2q^{-A_0}\right)U(\theta)\ket{x}_{\beta}\nl
&=\sum_{n=0}^{\infty}{}_{\beta}\bra{n}U(\theta)\ket{x}_{\beta}E_q^{1/2}\left(\theta^2\right)\left(-\theta^2q^{-n}\right)_n^{1/2}\sqrt{q^{\binom{n}{2}}\frac{\left(q^\beta;q\right)_n}{(q;q)_n}}t^n~.
\end{align}
This corresponds to a sum of $q$-Meixner polynomials. Indeed, using \eqref{eqXii}, we obtain
\begin{align}
\label{eqF1_1}
\mathcal{F}_1=E_q^{1/2}\left(\theta^2\right)(-\theta)^x{x+\beta-1 \brack x}^{\frac12}\sqrt{\frac{q^{\binom{x}{2}}}{\left(-\theta^2;q\right)_{x+\beta}}} \sum_{n=0}^{\infty}(\theta t)^n\frac{\left(q^\beta;q\right)_n}{(q;q)_n}\mathcal{M}_n\left(q^{-x};q^{\beta-1},\theta^2;q\right)~.
\end{align}
The other way to express $\mathcal{F}_1$ is to use \eqref{eqExp24} to write
\begin{align}
V(\theta,t)U(\theta)&=E_q\left(t(1-q)q^{\frac{B_0-A_0+1}{2}}A_-B_-\right)e_q\left(\theta(1-q)q^{\frac{B_0-A_0+1}{2}}A_+B_+\right)\nl
&\quad\times E_q\left(-\theta(1-q)q^{\frac{B_0-A_0+1}{2}}A_-B_-\right) E_q^{1/2}\left(\theta^2q^{B_0+1}\right)\nl
&=e_q\left(\theta tq^{-A_0}\right)e_q\left(\theta(1-q)q^{\frac{B_0-A_0+1}{2}}A_+B_+\right)E_q\left(t(1-q)q^{\frac{B_0-A_0+1}{2}}A_-B_-\right)\nl
&\quad\times E_q\left(-\theta tq^{B_0+1}\right)E_q\left(-\theta(1-q)q^{\frac{B_0-A_0+1}{2}}A_-B_-\right)E_q^{1/2}\left(\theta^2q^{B_0+1}\right)~,
\end{align}
which then leads to
\begin{align}
\hspace{-2em}\mathcal{F}_1&=e_q(\theta t)E_q^{1/2}\left(\theta^2q^{x+\beta}\right) \nl
&\times{}_{\beta}\bra{0}E_q\left(t(1-q)q^{\frac{B_0-A_0+1}{2}}A_-B_-\right)E_q\left(-\theta tq^{B_0+1}\right)E_q\left(-\theta(1-q)q^{\frac{B_0-A_0+1}{2}}A_-B_-\right)\ket{x}_{\beta}~.
\end{align}
Expanding the $q$-exponentials and recalling the orthogonality of the basis vectors, one obtains 
\begin{align}
\label{eqF1_2}
\mathcal{F}_1=\frac{E_q^{1/2}\left(\theta^2\right)}{\left(-\theta^2;q\right)_{x+\beta}^{1/2}}\frac{(-\theta)^x}{(\theta t;q)_\beta}{x+\beta-1 \brack x}_q^{\frac12}\sqrt{q^{\binom{x}{2}}}\ {}_{1}\phi_{1}\left({q^{-x}} \atop {\theta t q^\beta} \middle| q;-\frac{t q}{\theta} \right)~.
\end{align}
Performing the change of variable $z=\theta t$ and equating the RHS of \eqref{eqF1_1} and \eqref{eqF1_2} yield the following generating function with respect to the degrees :
\begin{align}
e_q(z)\,E_q\left(-zq^{\beta}\right){}_{1}\phi_{1}\left({q^{-x}} \atop {zq^\beta} \middle| q;-\frac{zq}{\theta^2} \right)=\sum_{n=0}^{\infty}z^n\frac{\left(q^\beta;q\right)_n}{(q;q)_n}\mathcal{M}_n\left(q^{-x};q^{\beta-1},\theta^2;q\right)~.
\end{align}
This generating function seems new to the best of our knowledge; it is also valid in general when $q^{\beta-1}$ is replaced by $b$ for $(0<b<1)$.

\subsection{Generating function with respect to the variables}
A generating function where the sum is over the variables is obtained in a fashion similar to what was done in the last subsection. Introduce
\begin{align}
W(\theta,t)=e_q^{1/2}\left(-\theta^2 q^{B_0+1}\right)e_q\left(t(1-q)q^{\frac{B_0-A_0+1}{2}}A_+B_+\right)
\end{align}
and let $\mathcal{F}_2={}_{\beta}\bra{n}U(\theta)W(\theta,t)\ket{0}_{\beta}$. From the definition of the $q$-exponentials \eqref{eqqExp}, one has
\begin{align}
\mathcal{F}_2&={}_{\beta}\bra{n}U(\theta)e_q^{1/2}\left(-\theta^2 q^{B_0+1}\right)e_q\left(t(1-q)q^{\frac{B_0-A_0+1}{2}}A_+B_+\right)\ket{0}_{\beta}\nl
&=\sum_{x=0}^\infty{}_{\beta}\bra{n}U(\theta)\ket{x}_{\beta}e_q^{1/2}\left(-\theta^2\right)\sqrt{\left(-\theta^2;q\right)_{x+\beta}q^{-\binom{x}{2}}}{x+\beta-1 \brack x}_q^{\frac12} t^x~.
\end{align}
In view of \eqref{eqXii}, one obtains
\begin{align}
\label{eqF2_1}
\hspace{-2em}\mathcal{F}_2&=e_q^{1/2}\!\left(-\theta^2\right)\sqrt{\frac{q^{-\binom{n}{2}}}{\left(-\theta^2q^{-n};q\right)_n}}{n+\beta-1 \brack n}_q^{\frac12}\!\!\!\theta^n \sum_{x=0}^{\infty}(-\theta t)^x{x+\beta-1 \brack x}_q\!\!\!\!\mathcal{M}_n\left(q^{-x};q^{\beta-1},\theta^2;q\right)~.
\end{align}
Meanwhile, using \eqref{eqExp24} we obtain
\begin{align}
U(\theta)&W(\theta,t)=e_q^{1/2}\left(-\theta^2q^{-A_0}\right) e_q\left(\theta(1-q)q^{\frac{B_0-A_0+1}{2}}A_+B_+\right)\nl
&\quad\times E_q\left(-\theta(1-q)q^{\frac{B_0-A_0+1}{2}}A_-B_-\right) e_q\left(t(1-q)q^{\frac{B_0-A_0+1}{2}}A_+B_+\right)\nl
&=e_q^{1/2}\left(-\theta^2q^{-A_0}\right)e_q\left(\theta(1-q)q^{\frac{B_0-A_0+1}{2}}A_+B_+\right)e_q\left(-\theta t q^{-A_0}\right)\nl
&=e_q\left(t(1-q)q^{\frac{B_0-A_0+1}{2}}A_+B_+\right)E_q\left(-\theta(1-q)q^{\frac{B_0-A_0+1}{2}}A_-B_-\right)E_q\left(\theta tq^{B_0+1}\right)~.
\end{align}
Then,
\begin{align}
\mathcal{F}_2&=e_q^{1/2}\left(-\theta^2q^{-n}\right)E_q\left( \theta tq^{\beta}\right)\nl
&\times {}_{\beta}\bra{n}e_q\left(\theta(1-q)q^{\frac{B_0-A_0+1}{2}}A_+B_+\right)e_q\left(-\theta t q^{-A_0}\right)e_q\left(t(1-q)q^{\frac{B_0-A_0+1}{2}}A_+B_+\right)\ket{0}_{\beta}~.
\end{align}
Expanding the $q$-exponentials and using the orthogonality of the basis vectors lead to
\begin{align}
\label{eqF2_2}
\mathcal{F}_2&=\frac{e_q^{1/2}\left(-\theta^2\right)}{\left(-\theta^2q^{-n};q\right)_n^{1/2}}\frac{\theta^n}{(-\theta t;q)_\beta}{n+\beta-1 \brack n}_q^{\frac12}\sqrt{q^{-\binom{n}{2}}}\ {}_{2}\phi_{1}\left({q^{-n},0} \atop {-\frac{q}{\theta t}} \middle| q;-\frac{q^{n+1}}{\theta^2} \right)~.
\end{align}
Effecting the change of variables $z=-\theta t$ and equating the RHS of \eqref{eqF2_1} and \eqref{eqF2_2} leads to
\begin{align}
\frac{1}{(z;q)_\beta}{}_{2}\phi_{1}\left({q^{-n},0} \atop {q/z} \middle| q;-\frac{q^{n+1}}{\theta^2} \right)=\sum_{x=0}^{\infty}z^x\frac{\left(q^\beta;q\right)_x}{(q;q)_x}\mathcal{M}_n\left(q^{-x};q^{\beta-1},\theta^2;q\right)~.
\end{align}

\section{Conclusion}
\label{sec:conclusion}
Summing up, we have provided an interpretation of the univariate $q$-Meixner polynomials where they arise as matrix elements of unitary $q$-pseudorotation representations on $q$-oscillator states. The unitarity of the representations was observed to imply the orthogonality relations for the polynomials and the structure relations were given a useful algebraic underpinning. A duality property has been presented and observed to lead to a new set of relations for the polynomials. Generating functions of two different types were obtained in this $\mathcal{U}_q(\mathfrak{su}(1,1))$ framework.

Now that this understanding of the univariate $q$-Meixner polynomials has been elaborated, it should be possible to develop an algebraic interpretation of the multivariate $q$-Meixner polynomials, 
similar to what was done for the multivariate $q$-Krawtchouk polynomials \cite{rf2016_Genest_Post_Vinet_algebqkraw}. It is expected that the $d$-dimensional $q$-Meixner polynomials will appear as matrix elements of unitary $(d+1)$-dimensional $q$-pseudorotation representations on $(d+1)$ $q$-oscillator states. The authors plan to report on this matter in future work.

\subsection*{Acnowledgments}
The authors would like to thank V. X. Genest, T. Koornwinder, M. E. H. Ismail, and A. Zhedanov for useful remarks and helpful discussions. J. G. holds an Alexander-Graham-Bell Graduate Scholarship from the Natural Sciences and Engineering Research Council (NSERC) of Canada. The research of L. V. was supported in part by the NSERC.

\appendix
\section{``Dual'' relations}
\label{app:AppendixA}
The standard relations derived in the paper can be given a dual version by the process explained in \autoref{sec:duality}. Here is the list of the relations that are obtained in this fashion.

Backward relation \eqref{eqBackRel} $\to$
\begin{align}
\theta^2q^{x+1}\left(1-q^\beta\right)\mathcal{M}_{n}\left(q^{-(x+1)};q^{\beta-1},\theta^2;q\right)&=\theta^2q^{x+1}\left(1-q^{n+\beta}\right)\mathcal{M}_{n}\left(q^{-x};q^{\beta},\theta^2;q\right)\nl
&\hspace{-6em}-q\left(1-q^n\right)\left(1+\theta^2q^{x+\beta}\right)\mathcal{M}_{n-1}\left(q^{-x};q^{\beta},\theta^2q^{-1};q\right)
\end{align}

Forward relation \eqref{eqForRel} $\to$
\begin{align}
\frac{1-q^{-x}}{\theta^2\left(1-q^\beta\right)}\mathcal{M}_{n}\left(q^{-(x-1)};q^{\beta-1},\theta^2;q\right)=\mathcal{M}_{n+1}\left(q^{-x};q^{\beta-1},\theta^2q;q\right)-\mathcal{M}_{n}\left(q^{-x};q^{\beta-1},\theta^2;q\right)
\end{align}

Difference equation \eqref{eqDiffRel} $\to$
\begin{align}
\left(1-q^x\right)\mathcal{M}_n\left(q^{-x};q^{\beta-1},\theta^2;q\right)&=-\left(1-q^n\right)\left(1+\theta^2q^{x+\beta-1}\right)\mathcal{M}_{n-1}\left(q^{-x};q^{\beta-1},\theta^2q^{-1};q\right)\nl
&\hspace{-4em}+\left[\left(1-q^n\right)\left(1+\theta^2q^{x+\beta-1}\right)+\theta^2q^x\left(1-q^{n+\beta}\right)\right]\mathcal{M}_n\left(q^{-x};q^{\beta-1},\theta^2;q\right)\nl
&\hspace{-4em}-\theta^2q^x\left(1-q^{n+\beta}\right)\mathcal{M}_{n+1}\left(q^{-x};q^{\beta-1},\theta^2q;q\right)
\end{align}

Complementary Backward relation \eqref{eqDualBack} $\to$
\begin{align}
\frac{q}{\theta^2}\frac{1-q^n}{1-q^{\beta-1}}\mathcal{M}_{n-1}\left(q^{-x};q^{\beta-1},\theta^2q^{-1};q\right)&=\mathcal{M}_{n}\left(q^{-x};q^{\beta-2},\theta^2;q\right)\nl
&-\mathcal{M}_{n}\left(q^{-(x+1)};q^{\beta-2},\theta^2q^{-1};q\right)
\end{align}

Complementary Forward relation \eqref{eqDualFor} $\to$
\begin{align}
\theta^2q^x\left(1-q^\beta\right)\mathcal{M}_{n+1}\left(q^{-x};q^{\beta-1},\theta^2q;q\right)&=\theta^2\left(1-q^{x+\beta}\right)\mathcal{M}_{n}\left(q^{-x};q^{\beta},\theta^2;q\right)\nl
&\hspace{-2em}-\left(q^n+\theta^2\right)\left(1-q^x\right)\mathcal{M}_{n}\left(q^{-(x-1)};q^{\beta},\theta^2q;q\right)
\end{align}

Recurrence relation \eqref{eqReccRel} $\to$
\begin{align}
q^{x+1}\left(1-q^{n}\right)\mathcal{M}_n\left(q^{-x};q^{\beta-1},\theta^2;q\right)&=-q\left(1-q^x\right)\left(q^n+\theta^2\right)\mathcal{M}_n\left(q^{-(x-1)};q^{\beta-1},\theta^2q;q\right)\nl
&\hspace{-4em}+\left[q\left(1-q^x\right)\left(q^n+\theta^2\right)+\theta^2\left(1-q^{x+\beta}\right) \right]\mathcal{M}_n\left(q^{-x};q^{\beta-1},\theta^2;q\right)\nl
&\hspace{-4em}-\theta^2\left(1-q^{x+\beta}\right)\mathcal{M}_n\left(q^{-(x+1)};q^{\beta-1},\theta^2q^{-1};q\right)
\end{align}

\section{Useful $q$-series identities}
\label{app:AppendixB}
A number of useful  $q$-series identities are gathered here for convenience.

The $q$-binomial coefficients are defined as follows
\begin{align}
\label{eqqBinom}
{n \brack k}_q =\frac{(q;q)_n}{(q;q)_k(q;q)_{n-k}},\qquad k=0,1,2,\ldots,n~.
\end{align}
They tend to the usual coefficients when $q\to 1$.

The little $q$-exponential, $e_q(z)$, and the big $q$-exponential, $E_q(z)$, are defined by
\begin{align}
\label{eqqExp}
e_q(z)={}_{1}\phi_{0}\left({0} \atop {-} \middle| q;z \right)=\frac{1}{(z;q)_\infty}~,\qquad 
E_q(z)={}_{0}\phi_{0}\left({-} \atop {-} \middle| q;-z \right)=(-z;q)_\infty~,
\end{align}
for $|z|<1$. It is straightforward to see that $e_q(z)E_q(-z)=1$. From \eqref{eqqExp}, one easily derives the following relations :
\begin{align}
\begin{aligned}
\label{eqIdeEqz}
&e_q(\lambda q^n)=e_q(\lambda)(\lambda;q)_n~,&\qquad &e_q(\lambda q^{-n})=\frac{e_q(\lambda)}{(\lambda q^{-n};q)_n}~,\\
&E_q(\lambda q^n)=\frac{E_q(\lambda)}{(-\lambda;q)_n}~,&\qquad &E_q(\lambda q^{-n})=E_q(\lambda)(-\lambda q^{-n};q)_n~.
\end{aligned}
\end{align}

The Baker–Campbell–Hausdorff formula admits two $q$-extensions \cite{rf1993_Floreanini_Vinet_automorphismqosc, rf1993_Kalnins_Miller_Mukherjee_matrixelemqoscalgebra}. The first one is
\begin{align}
\label{eqBCH1}
\begin{aligned}
&E_q(\lambda X)Ye_q(-\lambda q^\alpha X)=\sum_{n=0}^{\infty}\frac{\lambda^n}{(q;q)_n}[X,Y]_n~,\\
&[X,Y]_0=Y~,\qquad [X,Y]_{n+1}=q^nX[X,Y]_n-q^\alpha[X,Y]_nX~,\qquad n=0,1,2,\ldots
\end{aligned}
\end{align}
The second one is
\begin{align}
\label{eqBCH2}
\begin{aligned}
&e_q(\lambda X)YE_q(-\lambda q^\alpha X)=\sum_{n=0}^{\infty}\frac{\lambda^n}{(q;q)_n}[X,Y]_n^\prime~,\\
&[X,Y]_0^\prime=Y~,\qquad [X,Y]_{n+1}^\prime=X[X,Y]_n^\prime-q^{n+\alpha}[X,Y]_n^\prime X~,\qquad n=0,1,2,\ldots
\end{aligned}
\end{align}
Let us also record that for $XY=qYX$, one has
\begin{align}
\label{eqeqsum}
e_q(X+Y)=e_q(Y)e_q(X)~,\qquad \text{and}\qquad E_q(X+Y)=E_q(X)E_q(Y)~.
\end{align}


\section*{References}
\begin{thebibliography}{99}
\bibitem{rf2010_Koekoek_Lesky_Swarttouw_hypergeometric}
  R.~Koekoek, P.~A.~Lesky, and R.~F.~Swarttouw. 
  \newblock{\em Hypergeometric orthogonal polynomials and their $q$-analogues}.
  \newblock{Springer, $1^{\text{st}}$ edition, 2010}.
\bibitem{rf1934_Meixner_polyn}
  J.~Meixner. 
  \newblock{Orthogonale Polynomsysteme mit einer besonderen Gestalt der erzeugenden Funktion}.
  \newblock{\em J. London Math. Soc.}, s1-9:6–13, 1934.
\bibitem{rf1982_Basu_Wolf_irrepsSL2R}
  D.~Basu, and K.~B.~Wolf.
  \newblock{The unitary irreducible representations of $SL(2,R)$ in all subgroup reductions}.
  \newblock{\em Journal of Mathematical Physics}, 23:189-205, 1982.
\bibitem{rf1991_Vilenkin_Klimyk_specialfct}
  N.~Ja.~Vilenkin, and A.~U.~Klimyk.
  \newblock{\em Representation of Lie Groups and Special Functions}.
  \newblock{Springer, 1991}.
\bibitem{rf1986_Granovskii_Zhedanov_su11Meixner}
  Ya.~I.~Granovskii, and A.~S.~Zhedanov.
  \newblock{Orthogonal polynomials in the Lie algebras}.
  \newblock{\em Soviet Physics Journal}, 29(5):387-393, 1986. (transl. from Russian)
\bibitem{rf1993_Floreanini_LeTourneux_Vinet_quantmechpolyndiscrete}
  R.~Floreanini, J.~LeTourneux, and L.~Vinet.
  \newblock{Quantum mechanics and polynomials of a discrete variable}.
  \newblock{\em Annals of Physics}, 226(2):331-349, 1993.
\bibitem{rf1975_Griffiths_multivMeix}
  R.~C.~Griffiths.
  \newblock{Orthogonal polynomials on the negative multinomial distribution}.
  \newblock{\em Journal of Multivariate Analysis}, 5(2):271-277, 1975.
\bibitem{rf2014_Genest_Miki_Vinet_Zhedanov_JPhysA_multivMeixInterp}
  V.~X. Genest, H.~Miki, L.~Vinet, and A.~Zhedanov.
  \newblock {The multivariate Meixner polynomials as matrix elements of $SO(d, 1)$ representations on oscillator states}.
  \newblock {\em Journal of Physics A : Mathematical and Theoretical}, 47(4):045207, 2014.
\bibitem{rf1989_Koornwinder_repsuq2}
  T.~H.~Koornwinder.
  \newblock{Representations of the twisted SU(2) quantum group and some $q$-hypergeometric orthogonal polynomials}.
  \newblock{\em Indagationes Mathematicae (Proceedings)}, 92(1):97-117, 1989.
\bibitem{rf2000_Koelink_qkrawspherfct}
  E.~Koelink.
  \newblock{$q$-Krawtchouk polynomials as spherical functions on the Hecke algebra of type B}.
  \newblock{\em Transactions of the American Mathematical Society}, 352(10):4789-4813, 2000.
\bibitem{rf2004_Smirnov_Campigotto_qMeixAsDFctsu11}
  Y.~Smirnov, and C.~Campigotto.
  \newblock{The quantum $q$-Krawtchouk and $q$-Meixner polynomials and their related $D$-functions for the quantum group $SU_q(2)$ and $SU_q(1,1)$}.
  \newblock{\em Journal of Computational and Applied Mathematics}, 164:643-660, 2004.
\bibitem{rf2016_Genest_Post_Vinet_Yu_Zhedanov_qrotkraw}
  V.~X.~Genest, S.~Post, L.~Vinet, G.-F.~Yu, and A.~Zhedanov.
  \newblock{$q$-rotations and Krawtchouk polynomials}.
  \newblock{\em The Ramanujan Journal}, 40(2):335–357, 2016.
\bibitem{rf2016_Genest_Post_Vinet_algebqkraw}
  V.~X.~Genest, S.~Post, and L.~Vinet.
  \newblock{An algebraic interpretation of the multivariate $q$-Krawtchouk polynomials}.
  \newblock{\em The Ramanujan Journal}, 2016.
\bibitem{rf2007_Gasper_Rahman_multivqracah}
  G.~Gasper, and M.~Rahman.
  \newblock{Some systems of multivariable orthogonal $q$-Racah polynomials}
  \newblock{\em The Ramanujan Journal}, 13(1):389-405, 2007.
\bibitem{rf2003_Atakishiev_Atakishiev_Klimyk_qLaguerreqMeix}
  M.~N.~Atakishiev, N.~M.~Atakishiev, and A.~U.~Klimyk.
  \newblock{Big $q$-Laguerre and $q$-Meixner polynomials and representations of the quantum algebra $U_q(su_{1,1})$}.
  \newblock{\em Journal of Physics A: Mathematical and General}, 36(41):10335–10347, 2003.
\bibitem{rf2000_Rosengren_quantalgebAskWil}
  H.~ Rosengren.
  \newblock{A new quantum algebraic interpretation of the Askey-Wilson polynomials}.
  \newblock{\em Contemporary Mathematics}, 254:371-394, 2000.
\bibitem{rf1989_Macfarlane_qharmosc}
  A.~J.~Macfarlane.
  \newblock{On $q$-analogues of the quantum harmonic oscillator and the quantum group $SU(2)_q$}.
  \newblock{\em Journal of Physics A: Mathematical and General}, 22(21):4581-4588, 1989.
\bibitem{rf1989_Biedenharn_quantgroupandqboson}
  L.~C.~Biedenharn.
  \newblock{The quantum group $SU_q(2)$ and a $q$-analogue of the boson operators}.
  \newblock{\em Journal of Physics A: Mathematical and General}, 22(18):L873-L878, 1989.
\bibitem{rf1991_Floreanini_Vinet_qorthpolynoscillgroup}
  R.~Floreanini, and L.~Vinet.
  \newblock{$q$-Orthogonal polynomials and the oscillator quantum group}.
  \newblock{\em Letters in Mathematical Physics}, 22(1):45-54, 1991.
\bibitem{rf1993_Zhedanov_qrot}
  A.~Zhedanov.
  \newblock{Q rotations and other Q transformations as unitary nonlinear automorphisms of quantum algebras}.
  \newblock{\em Journal of Mathematical Physics}, 34(6):2631-2648, 1993.
\bibitem{rf1985_Truax_squeeze}
  D.~R.~Truax.
  \newblock{Baker-Cambpell-Hausdorff relations and unitarity of $SU(2)$ and $SU(1,1)$ squeeze operators}.
  \newblock{\em Physical Review D}, 31(8):1988-1991, 1985.
\bibitem{rf1993_Floreanini_Vinet_automorphismqosc}
  R.~Floreanini, and L.~Vinet.
  \newblock{Automorphisms of the $q$-oscillator algebra and basic orthogonal polynomials}.
  \newblock{\em Physics Letters A}, 180(6):393-401, 1993.
\bibitem{rf1993_Kalnins_Miller_Mukherjee_matrixelemqoscalgebra}
  E.~G.~Kalnins, W.~Miller, and S.~Mukherjee.
  \newblock{Models of $q$-algebra representations: Matrix elements of the $q$-oscillator algebra}.
  \newblock{\em Journal of Mathematical Physics}, 34:5333-5356, 1993.









\end{thebibliography}
\end{document}